\documentclass[11pt]{article}

\usepackage[T1]{fontenc}
\usepackage[utf8]{inputenc}
\usepackage{lmodern}

\usepackage[a4paper,margin=1in]{geometry}
\usepackage{setspace}
\usepackage{amsmath,amsfonts}
\usepackage{algorithmic}
\usepackage{algorithm}
\usepackage{array}
\usepackage{textcomp}
\usepackage{url}
\usepackage{verbatim}
\usepackage{graphicx}
\usepackage[colorlinks=true,
            linkcolor=black,
            citecolor=black,
            urlcolor=blue]{hyperref}
\usepackage{cleveref}
\usepackage{multicol}
\usepackage{booktabs}
\usepackage{orcidlink}
\usepackage{academicons}

\usepackage{subcaption}

\usepackage{natbib} % remove [numbers] if you want author-year via \citet/\citep only
\setcitestyle{authoryear,round,semicolon}

\usepackage{credits}
\credit{JS}{1,1,0,0,1,0,0,1,0,0,1,0,1,1}
\credit{AK}{0,0,1,0,0,0,0,0,0,0,1,1,1,1}
\credit{TN}{0,0,1,0,0,0,0,0,1,0,0,1,1,1}
\credit{KI}{1,0,0,1,0,1,0,0,0,0,0,0,1,1}
\credit{AM}{1,0,1,1,0,1,1,0,0,1,1,1,1,1}

\usepackage{ifthen}

\newboolean{anonymous}
\setboolean{anonymous}{FALSE} % Set to true for anonymous version, false for regular
\title{Early Prediction of Student Performance Using Bayesian Updating with Informative Priors Across Cohorts
}

\ifthenelse{\boolean{anonymous}}{
    \author{Anonymous author(s)}
}{
  \author{
    Jakob Schwerter\orcidlink{0000-0001-5818-2431}\\
    \small Hector Research Institute of Education Sciences and Psychology, University of T\"{u}bingen\\
    \small \texttt{jakob.schwerter@uni-tuebingen.de}
    \and
    Amer Krivo\v{s}ija\orcidlink{0000-0003-3088-6343}\\   
    \small Department of Statistics, TU Dortmund University\\
    \small \texttt{amer.krivosija@tu-dortmund.de}
    \and
    Tim Novak\\
    \small Department of Statistics, TU Dortmund University\\
    \small \texttt{tim.novak@tu-dortmund.de}
    \and
    Katja Ickstadt\\
    \small Department of Statistics, TU Dortmund University\\
    \small \texttt{katja.ickstadt@tu-dortmund.de}
    \and
    Alexander Munteanu\\
    \small Department of Statistics, TU Dortmund University\\
    \small \texttt{alexander.munteanu@tu-dortmund.de}
  }
}

\begin{document}
\pagenumbering{gobble} % no page numbers shown
\allowdisplaybreaks
\maketitle

\ifthenelse{\boolean{anonymous}}{}{
\begingroup
\small
\noindent\textbf{Declarations:}

\noindent\textbf{Author note.} We thank Rieke Deborah M\"{o}ller-Ehmcke for her support as a student assistant and for her contributions during the early stage of this project.
\par\medskip
\noindent\textbf{Funding.} The authors were supported by the project “From Prediction to Agile Interventions in the Social Sciences (FAIR)”
funded by the Ministry of Culture and Science (MKW.NRW), Germany and by
the TU Dortmund -- Center for Data Science and Simulation (DoDaS).

\noindent\textbf{Credit Statement.}
\insertcreditsstatement

\noindent\textbf{Ethics approval.}
This study re-uses data from another data collection which was ethical approved \citep{Schwerter.etal2026}. Only students who consent to participate in the study are included in the analysis.

\noindent\textbf{Availability of materials.}
Analysis code is available at \url{https://osf.io/bquzy/overview?view_only=82d5e70ec17a41cdb52330bddbdb9e3b}.

\par\bigskip
\endgroup
}

\newpage
\begin{abstract}
Early identification of at risk students in higher education depends on predictive models that maintain accuracy across successive cohorts---a requirement that single-cohort modeling approaches fail to meet. This study evaluates Bayesian updating with informative priors from a previous cohort to improve cross-cohort prediction robustness using digital trace data. We fit weekly Bayesian linear, logistic, and ordinal regression models with either uninformative default priors or informative priors derived from posterior distributions of a preceding cohort. Models were applied to six weekly self-regulated learning (SRL)-aligned engagement indicators from two consecutive cohorts of students in a blended first-year mathematics course ($N_1$ = 307; $N_2 = 323$). Outcomes were exam points, final grades, and a binary at risk indicator. The models were evaluated weekly based on accuracy, sensitivity, and RMSE. In the source cohort, performance was already substantial by week~6. In the target cohort, informative priors improved early classification: Logistic models with priors reduced misclassification by 22\% and false negatives by 38\% in week~3 relative to the uninformative default. Ordinal models with priors similarly showed the strongest improvements in early weeks, reducing misclassification by 42\% in week~2 and reaching an accuracy of .77 by week~4. Linear models showed little benefit from prior information. These findings demonstrate that Bayesian updating is a viable method for improving early classification performance across cohorts, with gains concentrated in the early weeks of the semester when current-cohort data are scarce.
%Early identification of students at risk of poor performance is central to intervention in introductory college courses. In this study, we tested whether iterative updating of Bayesian models could improve early prediction by carrying forward information from a prior cohort. Using digital trace data from two consecutive cohorts of students enrolled in a blended first-year mathematics course, we fit weekly Bayesian linear, logistic, and ordinal regression models to six indicators of engagement with embedded video questions and online exercises. Models for the second cohort were estimated with either default priors or informative priors derived from posterior distributions from the first cohort. Outcomes were exam points, final grades, and a binary indicator of risk for low performance. Classification models were more useful for early warning than linear score prediction. In the source cohort, predictive performance was already substantial by week 6. In the target cohort, informative priors improved early classification: Logistic models reduced misclassification by as much as 22.22\% and false negatives by as much as 38.46\% in week 3, and ordinal models reduced misclassification by as much as 42.5\% in week 2 and reached an accuracy of .7654 by week 4. Linear models showed little benefit from prior information. The findings suggest that Bayesian updating can strengthen early-warning systems when predictions are based on malleable engagement indicators rather than static background characteristics.
\end{abstract}


\newpage
\clearpage
\pagenumbering{arabic} % start normal numbering
\setcounter{page}{1}   % restart at page 1
\section{Introduction}

%Predictive models based on digital trace data are increasingly used to identify students at risk of poor academic performance. However, these models are typically developed within a single cohort and often exhibit reduced performance when applied to subsequent cohorts.

In technology-enhanced and blended learning environments, students continuously generate behavioral data through their interactions with digital platforms, providing a basis for data-informed early-warning systems \citep{Pardo.etal2017}. Studies have shown that the behavior captured in digital traces from online learning environments can classify students being at risk \citep{Arizmendi.etal2022, Bernacki.etal2020}. Thereby, an early classification of students at risk of underperforming is essential for enabling timely instructional support and targeted interventions in higher education. Without sufficiently early detection, the window for meaningful remediation narrows as the semester progresses, and students may already face outcomes that are difficult to reverse by the time poor performance becomes apparent on graded assessments \citep{Bernacki.etal2020, Arizmendi.etal2022}. %This is particularly relevant in first-year courses, where students struggle with the increased autonomy in higher education compared to high school \citep{Wolters.Brady2021} and must effectively regulate their own learning \citep{Schunk.Greene2018}. 

However, predictive models in education are typically developed and validated within a single cohort \citep{Arizmendi.etal2022}.
When models are transferred across cohorts, predictive performance has often declined \citep{Conijn.etal2017, Xing.etal2021}, suggesting that approaches which systematically integrate information from earlier iterations could improve robustness. Such declines likely reflect shifts in the relationship between behavioral features and performance outcomes across cohorts, for example due to changes in instructional design or student populations. Rather than treating these shifts as model failure, they can be understood as changes in the underlying data-generating process. What is needed, then, is a modeling approach that can formally incorporate information accumulated from earlier course offerings---stabilizing parameter estimates when current-cohort data are still scarce---while remaining flexible enough to adapt as new observations accumulate. Bayesian updating offers this capability: by encoding prior cohort results as informative prior distributions over model parameters, the framework carries forward empirical knowledge from previous iterations and revises it in light of incoming data from the current cohort \citep{Konig.VanDeSchoot2018}. 

Despite the potential of Bayesian updating to improve cross-cohort prediction, informative priors remain underused in educational research \citep{Konig.VanDeSchoot2018}. Educational studies rarely capitalize on posterior distributions learned from prior data collections, leaving available information on the table when new cohorts are analyzed.

%While predictive learning analytics has become a mature research domain, an important limitation remains. Predictive models are typically developed and evaluated within a single cohort and are often re-estimated independently for each course iteration, rather than cumulatively updated over time. 
%As a result, predictive performance frequently declines when models are applied to subsequent cohorts, particularly when instructional contexts or student populations shift \citep{Bernacki.etal2020, Arizmendi.etal2022, Xing.etal2021}. This limitation reflects the lack of approaches that systematically incorporate prior information to support model updating across cohorts. Bayesian statistics offers a principled framework to incorporate prior information into successive analyses, enabling continuous model updating and improving predictive robustness across iterations \citep{Konig.VanDeSchoot2018,gelman2013bayesian}. Despite this potential, prior information remains underused in educational Bayesian modeling \citep{Konig.VanDeSchoot2018}.

To address this limitation, we evaluate Bayesian regression models that incorporate information from previous cohorts through informative priors. The models use online behavioral traces as self-regulated learning (SRL) informed predictors and are evaluated in a week-by-week prediction framework. Thereby, this study makes three contributions. First, we evaluate the effect of incorporating informative priors from a previous cohort on predictive performance relative to models without prior information. Second, we introduce a week-by-week evaluation framework to analyze how predictive accuracy and sensitivity develop over time and to determine the earliest point at which reliable classification can be achieved. Third, we systematically compare Bayesian regression specifications and prior strengths to identify conditions under which prior information yields the greatest predictive benefit.

%Beyond the application context, this study provides a systematic evaluation of how prior information from previous datasets influences predictive performance over time in Bayesian regression models.

\section{Literature Review}

\subsection{Predictive Learning Analytics in Higher Education}

Predicting student performance has become a central focus of learning analytics research, with the goal of enabling timely instructional interventions and personalized feedback \citep{barshay_colleges_2019}. Numerous studies have explored predictive models using learning management system (LMS) data, in-class assessments, and learner characteristics to forecast academic outcomes \citep{Arizmendi.etal2022}. However, the predictive value of LMS features varies widely across courses, limiting model transferability and practical implementation \citep{Conijn.etal2017}. %Furthermore, many predictive models rely on static or non-malleable variables (e.g., demographics, prior GPA), which limits their usefulness for actionable early interventions.

Recent work advocates for the use of malleable behavioral features aligned with SRL theory, as they offer more interpretable and intervention-relevant insights \citep{Bernacki.etal2020}. Design-based work in learning analytics further suggests that prediction models grounded in theoretically meaningful behavioral indicators can produce insights that remain interpretable and actionable for instructors across course iterations \citep{plumley_codesigning_2024}. In a systematic review, \citet{Sghir.etal2023a} noted a trend toward more advanced machine learning techniques but highlighted a scarcity of approaches that remain interpretable for educators, especially regarding how predictions can guide pedagogical decision-making.

\subsection{Model Transferability and Continuous Knowledge Building}

A key challenge in predictive learning analytics is the limited generalizability of prediction model. Models trained on one cohort often yield reduced predictive performance when applied to subsequent course offerings or different course formats, due to changes in learner behavior, instructional design, or course structure \citep{Conijn.etal2017}. %Schwerter_generalizing_2026
\citet{Xing.etal2021} demonstrated that directly transferring predictive models to new course iterations can yield reduced performance, particularly when learner populations or course contexts shift between offerings. Bayesian updating techniques can incrementally integrate new data to improve transferability \citep{Konig.VanDeSchoot2018}. The findings of \citet{Xing.etal2021} suggest that predictive modeling should leverage information accumulated from previous cohorts rather than repeatedly building models from scratch.

%Despite the promise of cumulative modeling, the field has not yet fully embraced methods that support knowledge building across studies. \citet{Konig.VanDeSchoot2018} emphasized that educational research rarely makes use of informative priors, and as a result, does not capitalize on Bayesian methods for synthesizing evidence across multiple data collections. Addressing this gap may strengthen both the reproducibility and the cumulative validity of learning analytics research.

\subsection{SRL in Technology-Enhanced Settings}

SRL provides a theoretical lens for understanding student behaviors that relate to performance. SRL refers to the processes by which learners plan, monitor, and regulate their cognition, motivation, and behavior to achieve academic goals \citep{Zimmerman2002a,Schunk.Greene2018}. In digital learning environments, students’ interactions with online platforms provide observable indicators of (some of) these processes. By aligning these indicators with SRL frameworks, engagement behaviors such as attempting self-assessment activities or managing study time relative to deadlines can be linked to regulatory processes described in models of SRL, including planning, monitoring, and strategy use \citep{Du.etal2023b, Bernacki.etal2020, bernacki_using_2025}. Recent multimodal research shows that digital trace data can approximate SRL-related behaviors \citep{bernacki_using_2025, Du.etal2023b, Colling.etal2025} and support performance prediction in higher education \citep{Bernacki.etal2020,cogliano2022self}. %Sequence-based analyses of learning events further show that patterns of engagement with instructional materials can reveal SRL strategies associated with higher achievement \citep{yu_interpreting_2025, Colling.etal2025}.

\subsection{Summary and Research Gap}

Existing research highlights a central limitation that motivates the present study: predictive learning analytics models are typically not updated cumulatively across cohorts, limiting their robustness under changing instructional conditions or shifts in student populations. %Although SRL-informed digital trace features provide theoretically meaningful and interpretable predictors, their predictive value may vary across contexts and over time. 
This variability makes it especially important to adopt modeling approaches that can leverage stable signal from prior cohorts while accommodating such shifts in the current one. Building on this, the present study evaluates whether Bayesian regression models with informative priors can improve early prediction and robustness across cohorts, while also assessing prediction performance in a week-by-week framework.
Therefore, this study is guided by the following two research questions: %\textcolor{red}{Moved the questions here}
%\section{Research Questions}
\begin{enumerate}
    \itemsep-2pt 
    %    \item How accurately can Bayesian regression models predict student performance at early stages of a blended first-year mathematics course?
    \item At what point during the semester do Bayesian regression models achieve reliable predictive performance in the source cohort?
    
    \item Does incorporating informative priors from a previous cohort improve predictive performance compared to using uninformative default priors and if so, how much earlier do we achieve reliable predictive performance? Furthermore, does this benefit differ across linear, logistic, and ordinal regression models?
    
    %\item Which Bayesian regression approach (linear, logistic, ordinal) and which prior specification yield the most reliable early-warning performance?
 \end{enumerate}

\section{Methodology}

\subsection{Context, Data, and Cohorts}
%\setboolean{anonymous}{TRUE}
The data used in this study were collected from the project\ifthenelse{\boolean{anonymous}}{
    % Anonymous version: omit author details
    [blinded for review].
    %\date{}
}{
  ``From Prediction to Agile Interventions in the Social Sciences (FAIR)'' \citep{Schwerter.etal2026}.
} 
Participants were undergraduate students enrolled in an introductory gateway mathematics course for business administration majors at a large public university in the south of Germany. The course is compulsory and is usually taken in the first semester of the Bachelor’s program. The study used data from two consecutive cohorts enrolled in the winter terms of 2022/2023 (hereafter WS1), and 2023/2024 (hereafter WS2). Both courses were taught by the same lecturer, and used the same (online) materials and grading standards. The only substantive difference between the two cohorts was therefore the student composition. 

%\textcolor{cyan}{Note of Tim:  Respectively for (WS1, WS2): total students enrolled (636, 584), no registration for exam (315, 249), not attended exam (14, 12), attended and failed exam (78, 80), passed (229, 243); mentioning also that we removed all students that did not register for or attended the exam ergo models trained with filtered students (307, 323)}
%\textcolor{cyan}{Note of Tim: Where do we explain the train-test splits? Here or in start of III C? 75\%-25\% ratio}

The course followed a blended learning format. Weekly in-person lectures were accompanied by weekly (offline) worksheets. The solutions to these worksheets were presented online in tutorial videos. These videos embedded multiple-choice questions at key steps of the worked out examples to prevent mind wandering and to give students the opportunity to test themselves, triggering the test-enhanced learning effect \citep{Schwerter.etal2026, roediger_test-enhanced_2006}. To provide students with an additional practice opportunity in the week thereafter, parameterized online exercises were generated from the worksheets and available for one week.
As is typical for large lecture courses in German higher education, attendance in both cohorts was not mandatory, and enrollment did not necessarily imply active participation or intent to complete the final exam.

The total number of enrolled students participating in this study was 636 in WS1 and 584 in WS2. 307 students in WS1 and 323 students in WS2 registered for the first of two dates of the final exam and attended the exam, of whom 78 in WS1 and 80 in WS2 did not pass. The present study focuses on students who sat the exam on the first exam date, which fell two weeks after the end of the lecture period. Students who registered for the second exam date---scheduled one week before the start of the following semester---were excluded from the analytic sample for two reasons. First, the extended timeline between the end of the lecture period and the second sitting means that the week-by-week prediction framework developed here is not directly applicable to this group, as their engagement trajectory extends beyond the observation window used for modeling. Second, pooling students across two exam dates with different timelines would introduce heterogeneous outcome timing that complicates model evaluation. We acknowledge that students who registered for the second exam date may differ systematically from those who sat the first — for instance, they may be disproportionately at risk of underperforming — and this restriction should be considered when interpreting the generalizability of the findings.

\subsection{Feature and Outcome Variables}%\textcolor{red}{I think measures refers to evaluation metrics rather than variables (but I might be wrong). Anyway, we need to describe variables but also test/train split and acc/sens and other measures.}

The analysis focused on learners’ weekly engagement with the digital learning materials as feature variables and their performance in the course as outcome variables. All features used to build the predictive models were selected based on availability for both cohorts and theoretical relevance to early performance monitoring \citep{Du.etal2023b, matcha2020analytics, yu_interpreting_2025}. %Schwerter.etal2026, 

\subsubsection{Engagement and Performance Indicators}

Six weekly indicators were extracted from the learning platform logs and the online exercise system. The indicators were recorded on a weekly basis to enable early and reliable prediction throughout the semester:
\begin{itemize}
    \itemsep-2pt
    \item Ratio of answered multiple-choice questions in the weekly interactive solution videos (answered ratio). %\textbf{Beantwortet:} 
    \item Ratio of correctly answered multiple-choice questions in the weekly solution videos (correct ratio). %\textbf{Korrekt:} 
    \item Days between submission and deadline of the weekly online exercise (first access to deadline).
    \item Ratio of correctly solved problems within the weekly online exercises (points ratio).% sheet that the student attempted. %\textbf{punkte.ratio:} 
    \item Total time spent working on the weekly online exercises (total time). %(sum of time spent across all problems in a given week). %\textbf{Bearbeitungsdauer:}
    \item Ratio of attempted problems within the weekly online exercises (worked-on ratio).% that the student attempted. %\textbf{bearbeitet.ratio:} 
\end{itemize}

\begin{figure}[htbp]
    \centering
    \includegraphics[width=\linewidth]{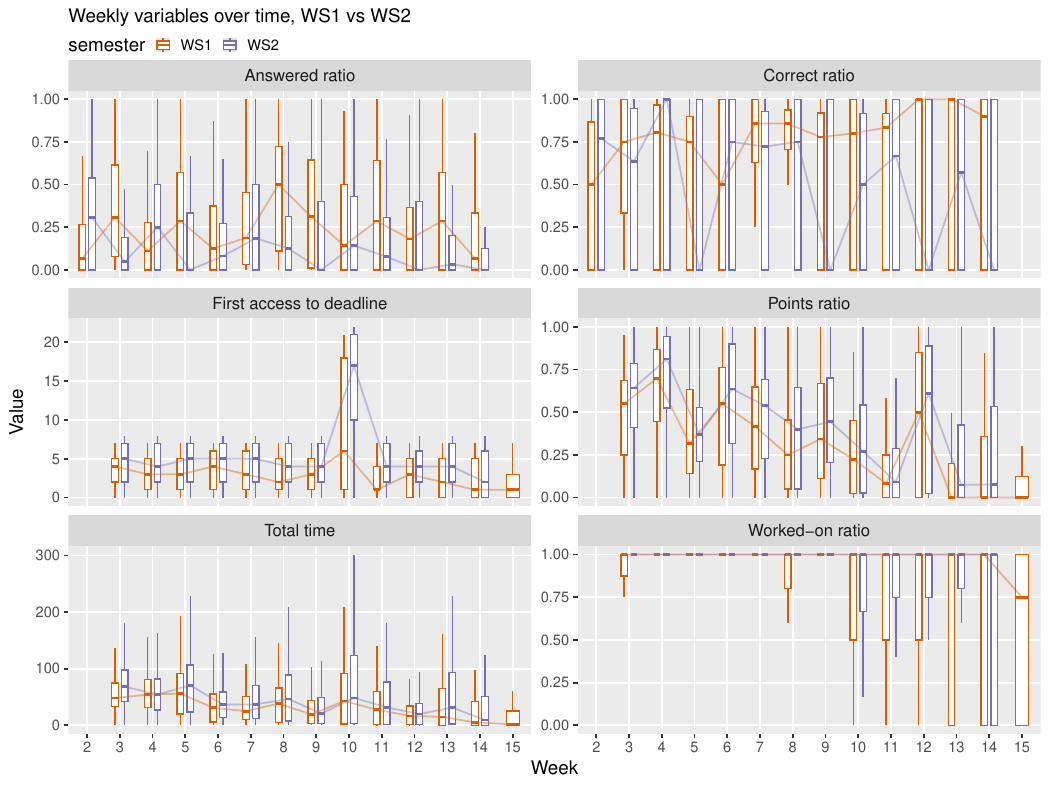}
    \caption{Weekly trajectories of the six engagement indicators for both cohorts (WS1: winter semester 2022/2023; WS2: winter semester 2023/2024) from week 2 through week 15. Each panel corresponds to one indicator. Boxes represent the interquartile range across students within each week; horizontal lines within boxes indicate medians.}
    \label{fig:weekly_development}
\end{figure}

\Cref{fig:weekly_development} displays the weekly trajectories of all six engagement indicators across both cohorts (WS1, WS2), illustrating the within-semester development of each variable from week 2 through week 15. The broadly comparable distributions of WS1 and WS2 across most indicators and weeks support the assumption that behavioral engagement patterns were sufficiently stable across cohorts to justify using posterior distributions from WS1 as informative priors for WS2. Two features of the figure reflect course-structural differences rather than behavioral patterns. First, the first access to deadline indicator is notably elevated in week 10 in both cohorts, because the Christmas break extended the interval between students' first access to the exercise and the submission deadline by approximately three weeks. Second, WS1 includes an additional online exercise in week 15 that was not offered in WS2. Neither feature affects the analyses reported here, as the target cohort (WS2) is analyzed only through week 10 (see Section~\ref{sec:bayesianupdate}).

The variables were not accumulated across weeks. Instead, they were appended to the weekly models as soon as they became available. For instance, the model fitted in week~5 comprised information on all weeks~$1-5$ of the current semester for each of the aforementioned variables. Information on weeks~$6-15$, lying in the future, were not (yet) included in the model, to simulate a realistic model development over time.

Missing features were always imputed by zero values. Zero imputations are semantically appropriate for all selected feature variables since missing values indicate a lack of students' engagement with the corresponding practice material. Their ratios of (correctly) answered or solved tasks can thus be reasonably quantified by zero. Similarly, their time spent working on the material is zero as well as their remaining time of submission, where the value zero represents a maximum amount of procrastination until the deadline passed.

\subsubsection{Outcome Variables}

The primary outcome was a binary indicator representing whether the students were ``at risk'' of underperforming, enabling the modeling of early-risk detection. The threshold was chosen to be at 3.3, which corresponds to a grade that indicates distinctly below average performance on the German grading scale, comparable to C- in US letter grades. This threshold is a more sensible choice than the pass/fail boundary, since it includes a margin comprising students who would possibly pass the exam by a small margin with undesirable grades and who could equally benefit from early detection helping to improve their SRL behavior and final performance. The threshold 3.3 also subdivides the population into two groups of roughly equal proportions: in the first cohort WS1, 158 (51.47\%) out of 307 students were considered at risk and the remaining 149 (48.53\%) students were not at risk. In the second cohort WS2, 166 (51.39\%) out of 323 students were considered at risk and the remaining 157 (48.61\%) students were not at risk.

As secondary outcome variables, we considered the final exam achievements measured either in scored points, i.e., integers between $0-83$ or grades on the German grade scale $(1.0, 1.3, 1.7, 2.0,\ldots,3.7,4.0,5.0)$, encoded numerically by integers $1-11$, where $5.0$ resp.\ $11$ represent `fail'.

\subsection{Regression Models}\label{sec:regression_models}
To predict the outcome variables $y$ based on the available feature variables per week in a vector of predictors $x$, we fitted the main model parameters $\beta$ of different Bayesian regression models, depending on the type of the outcome variable $y$.

\subsubsection{Linear Regression for Numerical Data}
To predict the exam scores as measured in terms of achieved points between $0-83$, the scale is sufficiently wide and fine-grained so it can be regarded as an approximately continuous variable that can be modeled using linear regression. In linear regression, the outcome variable is modeled using a linear relationship
\[
    y_i = x_i^T\beta + \epsilon
    %\hat \beta = \underset{\beta}{\argmin} \sum_{i=1}^n (y_i - x_i^T\beta)^2
\]
with Gaussian noise $\epsilon\sim\mathcal{N}(0,\varsigma^2)$, which implies that $p(y_i\mid\beta)$ follows a Gaussian distribution as well. The model variance $\varsigma^2$ is usually unknown but can be estimated within the Bayesian analysis framework.
Predictions are generated by evaluating the linear predictor without noise, $\hat y_i = x_i^T\beta$, which corresponds to the mean of $p(y_i\mid\beta)$.
%that minimize the sum of squared deviation of the $n$ observed values from the fitted linear prediction.

\subsubsection{Logistic Regression for Dichotomous Data}
To predict the probability of a student for being at risk (final grade $\geq 3.3$), logistic regression was employed, where the binary outcome is linked to the linear predictor using the `logit' link
\[
    p({y_i}\mid\beta) = \frac{1}{1+\exp(-x_i^T\beta)}\,.
\]
Classification can be performed by thresholding the probability at $.5$, i.e.,
\[
    \hat y_i = \begin{cases}
    1, \text{ if } p({y_i}) \geq .5\\
    0, \text{ if } p({y_i}) < .5\,.
    \end{cases}
\]

\subsubsection{Ordinal Regression for Ordered Categorical Data}
\allowdisplaybreaks
Ordinal regression extends the logit model of a binary logistic regression to the case of multiple categories that obey a natural ordering. In our case, the categories correspond to $11$ German grade scale levels encoded by integers 1 to 11. In addition to the linear parameters $\beta$, class thresholds $\theta_j$, to be used in the logit link, are estimated so that the probability for class $j$ is calculated as $p(y_i = j \mid \beta) = p(y_i \leq j \mid \beta)-p(y_i \leq j-1 \mid \beta)$, where
\[
    p(y_i \leq j \mid \beta) = \frac{1}{1+\exp(-(\theta_j-x_i^T\beta))}\,.
\]
% The probability for class $j$ can be derived as
% \[
%     p(y_i = j \mid \beta) = p(y_i \leq j \mid \beta)-p(y_i \leq j-1 \mid \beta)\,.
% \]
Classification into predicted grade levels $\hat y_i$ can then be performed by comparing the linear predictor to the calculated thresholds, i.e.,
\[
    \hat y_i = \begin{cases}
    1, \text{ if } x_i^T\beta \leq \theta_1\\
    2, \text{ if } \theta_1 < x_i^T\beta \leq \theta_2\\
    3, \text{ if } \theta_2 < x_i^T\beta \leq \theta_3\\
    \ldots\\
    11, \text{ if } \theta_{10} < x_i^T\beta\,.
    \end{cases}
\]

\subsection{Evaluation Measures}

The data of each semester was subdivided into a 75\%/25\% stratified train/test split for model selection and evaluation.
We used three different evaluation measures for assessing the performance of the predictive models. For the evaluation of predicted numerical outcomes such as exam scores, we used the root mean squared error (RMSE)
\[
    \text{RMSE} = \sqrt{\frac{1}{n}\sum\nolimits_{i=1}^n(\hat y_i-y_i)^2}\,,
\]
where $n$ is the size of the test dataset.
It can be interpreted as an estimate of the standard deviation of the predictions $\hat y_i$ from their true observed values $y_i$.

For classification, we used the accuracy (ACC), which measures the ratio of correctly classified observations among all observations of the test dataset. In the case of ordinal regression, a prediction was considered correct, if it lay within an interval of plus or minus $k=3$ around the true grade level. This criterion was introduced to allow for small deviations while still reflecting whether the predicted grade was within a reasonably narrow range around the true outcome.

For binary classification, we additionally used the sensitivity (SENS) as evaluation measure. It measures the ratio of students that are identified correctly for being at risk among all students who are actually at risk (final grade $\geq 3.3$). This measure is often more important than the accuracy in our context, because predicting a student to be at risk even if they do not fall in this category is less problematic than missing out on identifying a student who is actually at risk of underperforming.

\subsection{Bayesian Estimation and Updating Across Cohorts}\label{sec:bayesianupdate}

In contrast to traditional frequentist estimation, Bayesian statistics not only seeks for a single parameter vector $\beta$ that best fits the data. Instead, one seeks to sample from a posterior distribution $\pi_{\rm post}(\beta)$ over parameter vectors. This is done by incorporating knowledge from the present data in terms of a likelihood $p(y\mid\beta)$ under one of the regression models of interest, introduced in \Cref{sec:regression_models}. Another distribution $\pi_{\rm pre}(\beta)$ integrates prior belief on the parameters. Using Bayes' rule, the posterior can be expressed as a compromise between these components
\begin{equation}\label{eq:bayes}
    \pi_{\rm post}(\beta) \propto p(y\mid\beta)\cdot \pi_{\rm pre}(\beta)\,.
\end{equation}
In case no prior information is available, one may choose an improper uniform distribution over the parameter space, which is neutral with respect to \Cref{eq:bayes} and results in a proper posterior distribution as long as the likelihood is non-degenerate. We note that there exist other model-specific weakly-informative default priors, like Cauchy, Student-$t$ or Laplace distributions \citep{gelman2008weakly}.
However, for the main model parameters $\beta_j$, we choose to use the aforementioned non-informative prior and denote it the `default prior' which acts as a baseline to compare to. This is a sensible choice in our context, as it represents that \emph{actually no} prior information is available and thus allows to isolate the effect of incorporating informative prior information as compared to the default prior.

For secondary model parameters, such as the model variance $\varsigma^2$ in the linear regression model, and the threshold parameters $\theta_j$ in the ordinal model, the $\texttt{brms}$ package \citep{Buerkner2017}, that was employed for the analysis, uses default priors, like half-Cauchy or Student-$t$, based on the established standards in Bayesian statistics \citep{gelman2013bayesian}. We thus left the prior choice to the $\texttt{brms}$ package's default settings, as there were no indications that required alternative prior specification.

After performing the Bayesian analysis on one dataset, we gain a posterior distribution on the parameters given the observed data. One unique strength of Bayesian analyses is that this posterior distribution can be updated when new data is observed. 
To this end, the posterior of the previous analysis is incorporated as a prior distribution into the next Bayesian analysis. The new data is incorporated in form of the likelihood distribution. This results in the updated posterior distribution given the new data and the prior belief (carrying information obtained from previous data analyses), see \Cref{eq:bayes}.

This iterative approach addresses a central challenge in educational modeling: the instability of the data-generating process across different student cohorts. By using informative priors in subsequent analyses, we regularize the model against cohort-specific noise---such as idiosyncratic fluctuations in initial course motivation or other student characteristics---while retaining the signal from stable behavioral patterns identified in the first cohort. This approach should yield better predictive models than estimating each cohort from scratch with uninformative default priors.

In this study, linear, logistic and ordinal Bayesian regression models were fitted and their weekly ability of predicting the students of being at risk was evaluated. To investigate whether Bayesian modeling effectively benefits from prior cohort information, the first cohort (WS1, 2022/2023) served as the \textit{source cohort} for estimating posterior parameter distributions using the non-informative default prior distribution. The posterior estimates were then used as informative priors for the analysis of the \textit{target cohort} (WS2, 2023/2024), allowing a comparison between models with uninformative default prior and those incorporating informative prior information from the earlier cohort.

\begin{figure}[!htb]%[!ht]
\centering
\includegraphics[width=0.47\textwidth]{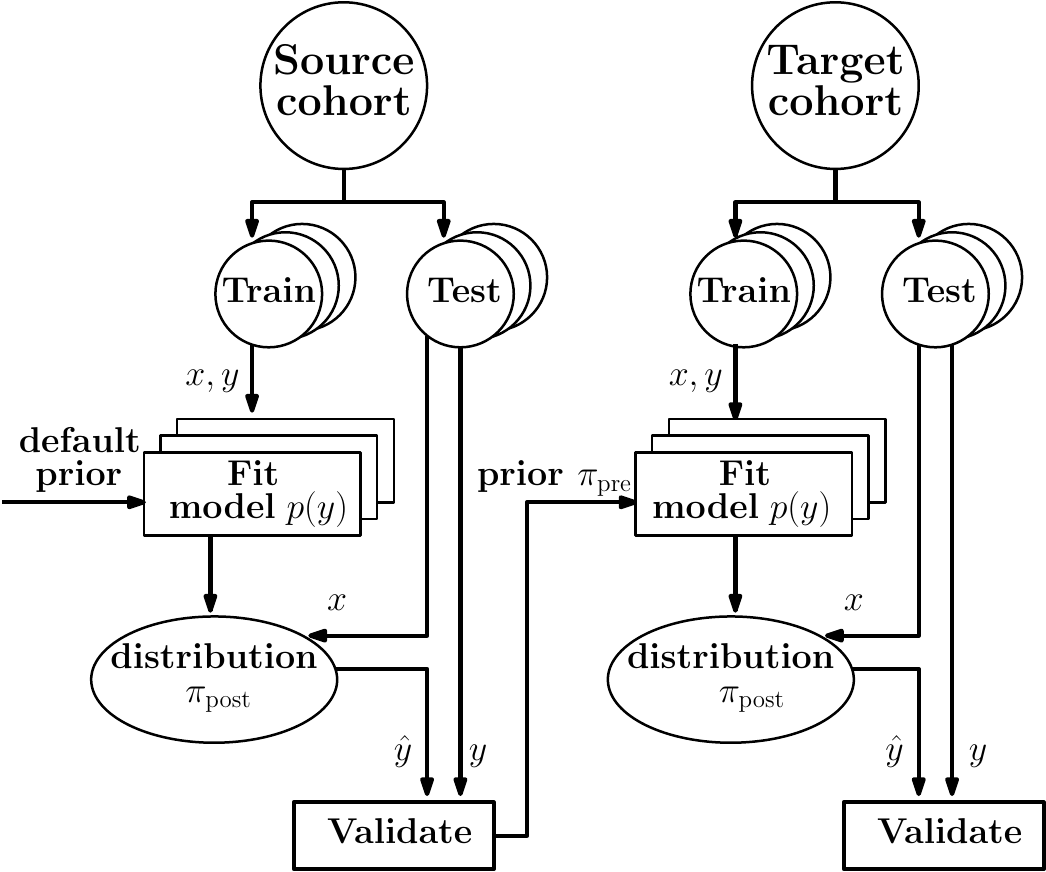}
\caption{Illustration of the Bayesian workflow, see\ \Cref{eq:bayes}.}
\label{fig:bayes}
\end{figure}

\Cref{fig:bayes} illustrates the Bayesian update procedure on the two subsequent cohorts in this study. For the {source cohort}, Bayesian regression models were first estimated without incorporating prior information (i.e., using the default prior). Based on these results, one week was selected for each model type according to their best predictive performance on test data.

The corresponding model was then refitted for the selected week using all available data, i.e., the combined training and test data. The resulting posterior parameter distributions from this refitted model served as informative priors for the target cohort. By additionally varying prior strength through scaling their standard deviations, this approach enabled a direct comparison between models with different prior strengths and without prior information.

More specifically, to investigate the effect of including prior information, two prior specifications were compared for the target cohort models: (1) uninformative default priors, and (2) normal priors centered at the posterior means of the source cohort, with their standard deviations scaled by multiplicative factors $\eta$ specified below. Means and standard deviations were estimated from parameter samples using standard maximum likelihood estimation. I.e., the prior information for each parameter $\beta_i$ was modeled as
\begin{align*}
    \beta_i &\sim \mathcal N(\mu_i, \eta\cdot\sigma_i)\,, \\[5pt]
    \mu_i &= \frac{1}{N}\sum\nolimits_{j=1}^N s_j\,, \\
    \sigma_i &= \sqrt{\frac{1}{N}\sum\nolimits_{j=1}^N (s_j-\mu_i)^2}\,,
\end{align*}
where $N$ denotes the number of parameter samples $s_j$ from the preceding posterior estimation.

Following an exploratory evaluation of prior strength, their standard deviations were multiplied by selected representative factors $\eta \in\{.5, 1, 1.5, 3, 6\}$, to determine the influence of more or less informative priors. This procedure enabled an examination of how strongly prior information should be weighted to optimize predictive performance especially in early weeks of the semester.

Weekly models were estimated based on the six engagement variables for each respective week. Comparisons between models with and without informative priors were conducted separately for each regression type and each week, to identify the point in the semester at which prior information most effectively improved prediction quality.

The predictive models were intended for an early warning of students being at risk, and for allowing to apply targeted intervention sufficiently early to unfold an effect. Therefore, data of the second cohort (WS2) were cut at week~10 (which was also the week of the Christmas break), because after this point, only 5 weeks of instruction remain before the final exams.

All Bayesian models were estimated using the \texttt{brms} package \citep{Buerkner2017} in \texttt{R} \citep{R2025}, which employs a Hamiltonian Monte Carlo \citep{hoffman2014no} sampler, a state of the art sampling method for Bayesian models. Posterior predictive distributions were used to derive weekly predictions, propagating parameter uncertainty to the generated predicted values for new or test observations, respectively. To assess predictive uncertainty, all predictions were repeated $1\,001$ times, to allow visualizing the median and quantiles of predictive distributions using boxplots. Reported numbers refer to the median unless explicitly stated otherwise.

\section{Results}

All reported analyses followed the Bayesian regression framework described in \Cref{sec:bayesianupdate} to evaluate the extent to which weekly indicators of student engagement predicted end-of-term performance, and whether incorporating prior information from the previous cohort improved predictive accuracy. Separate models were estimated for each week including the variables of all preceding weeks of the semester to assess how early reliable predictions could be achieved based on the full information up to the current week but without information on the future.

\subsection{Source Cohort: Prediction Stability Across Weeks (RQ1)}

Bayesian models were first trained for every week on the source cohort without prior information to generate default posterior distributions for each regression type (linear, logistic, ordinal). To ensure convergence in model fitting, the total number of iterations was increased from the default to 10\,000 per Hamiltonian Monte Carlo chain. %There were no indications of problematic behavior. Therefore, 
All remaining settings of the \texttt{brms} package were left at their defaults \citep[cf.][]{Buerkner2017}, as discussed in \Cref{sec:bayesianupdate}.

\begin{figure}[htbp]
    \centering
    \includegraphics[width=.675\linewidth]{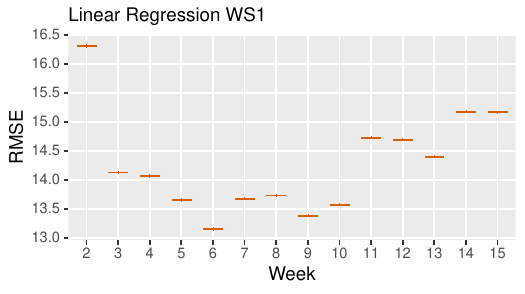}
    \caption{RMSE of the linear regression on the source cohort WS1. Lower is better. Best result in week~6. }
    \label{fig:lin:source}
\end{figure}

\begin{figure}[htbp]
    \centering
    \includegraphics[width=.675\linewidth]{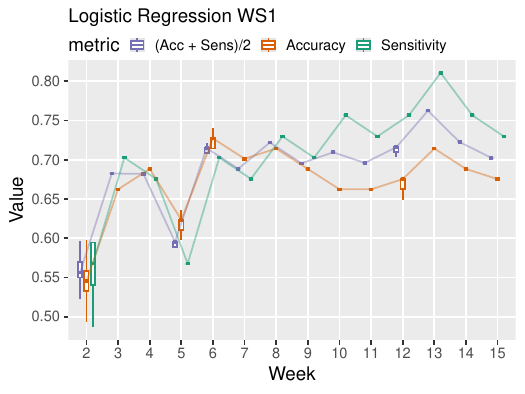}
    \caption{Accuracy and sensitivity of the logistic regression on the source cohort WS1. Higher is better. Best result in week~13.}
    \label{fig:log:source}
\end{figure}

\begin{figure}[htbp]
    \centering
    \includegraphics[width=.675\linewidth]{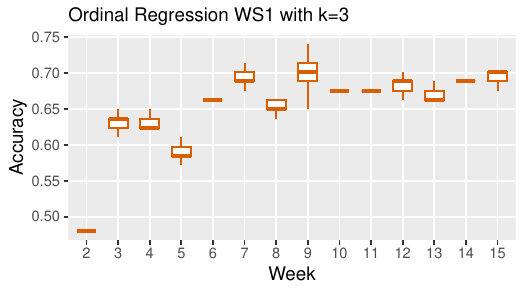}
    \caption{Accuracy of the ordinal regression on the source cohort WS1. Higher is better. Best result in weeks~9 and 15.}
    \label{fig:ord:source}
\end{figure}

For the source cohort, the predictive performance was already reasonably strong after 6 weeks with an accuracy and sensitivity above .7 (see Figures~\ref{fig:lin:source}, \ref{fig:log:source} and \ref{fig:ord:source}). After week~6, accuracies plateaued above $.66$ (logistic), $.65$ (ordinal), and RMSEs for the linear model remained above $13.3$ points (on the 0–83 point scale), indicating no substantial improvement beyond that point.

Only for the sensitivity in case of logistic regression, we observed considerable gains in later weeks at the cost of a decline in accuracy (see \Cref{fig:log:source}). On average, no substantial gain in predictive performance was observed after week~6, suggesting that the engagement indicators available by that point already carried most of the predictive signal for end-of-term outcomes. %This made week~6 a plausible earliest time instant at which predictions from the source cohort could inform intervention decisions.

\subsection{Selection of Prior Distributions for the Target Cohort}

While week~6 marks the point at which source cohort predictions became sufficiently accurate to support intervention decisions, a separate question concerns which source cohort model should supply the informative priors for the target cohort. Any week of the source cohort may be considered as a source of prior information. In our context it is preferable to choose a week that shows highest possible predictive performance, while it appears late in the semester to allow for propagating prior information on as many weeks as possible. This is important, since the model in a specific week includes only information on preceding weeks, not on subsequent weeks.

This motivates the prior to be taken from the empirically best-performing week for each model type. The best performing weeks usually appeared sufficiently late in the semester to allow for full prior specification in subsequent analyses. Following the procedure described above, the best-performing model per regression type in the source cohort was as follows. (i) For linear regression, the best predictive performance was achieved in week~6 (see \Cref{fig:lin:source}) with an RMSE of $13.15$. (ii) For logistic regression, the best predictive performance averaged over the two measures accuracy and sensitivity was at $.76$, %$.7625$
observed in week~13 (see \Cref{fig:log:source}). In particular the more important sensitivity measure reached its peak at $.81$ %$.8108$
in week~13. The accuracy in week~13 was at $.71$, %$.7143$
slightly below the peak of $.73$ %$.7273$
attained in week~6. Although week~6 represented the earliest point at which predictions were sufficiently accurate to support intervention decisions in the source cohort, week~13 was selected as the source of informative priors because it offered both the highest overall predictive performance and prior information on the largest number of weekly feature variables---a distinction between the intervention threshold and the optimal prior derivation point. (iii) Lastly, for ordinal regression, the best predictive performance was equally attained in weeks~9 and 15 (see \Cref{fig:ord:source}), each reaching an accuracy of $.7$. %$.7013$
Week~15 was selected because, at equal accuracy, the later week provides prior information on a larger number of weekly feature variables for their subsequent use in the target cohort.

Because each model type's priors were derived from a specific week, informative prior distributions were available only for predictors corresponding to weeks up to and including that selected week. For each model type, any predictor corresponding to a later week received an uninformative default prior instead. In practice, this distinction mattered only for linear regression, where the best-performing week was week~6: predictors for weeks 7–10 in the target cohort received default priors. For logistic and ordinal regression, the selected weeks (13 and 15, respectively) were preceded by the week-10 cutoff applied in the target cohort, so all predictors received informative priors.

\subsection{Target Cohort: Effect of Informative Priors on Early Classification (RQ2)} %(Addresses RQ2 — do priors improve performance?) But this is intertwined with RQ3?
%For the target cohort, only models up to week~10 were considered, justified by the assumption that an early intervention should be applied no later than the Christmas break in winter term, because after this point, only 5 weeks of instruction remain before the final exams. This reflects the latest practically relevant time point at which instructors could apply targeted intervention sufficiently early to unfold an effect.

\begin{figure}[htbp]
    \centering
    \includegraphics[width=.7\linewidth]{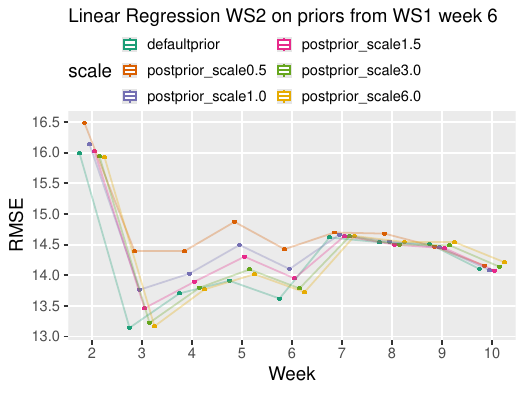}
    \caption{RMSE of the linear regression on the target cohort WS2 with the best prior from WS1 (see \Cref{fig:lin:source}) and with non-informative default prior. Standard deviations of informative priors are scaled by factors $.5$, $1.0$, $1.5$, $3.0$, and $6.0$.}
    \label{fig:lin:posterior}
\end{figure}

\begin{figure}[htbp]
    \centering
    \includegraphics[width=\linewidth]%,height=.45\linewidth]
    {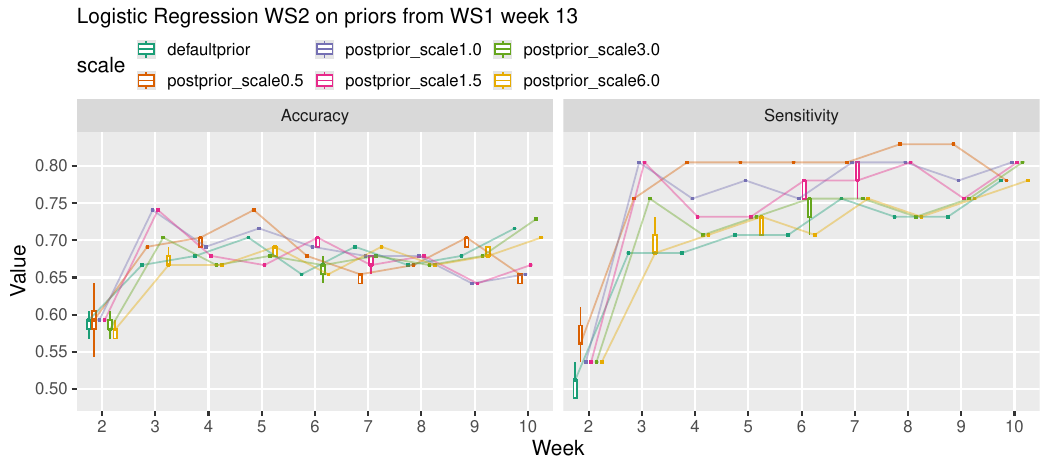}
    \caption{Accuracy and sensitivity of the logistic regression on the target cohort WS2 with the best prior from WS1 (see \Cref{fig:log:source}) and with non-informative default prior. Standard deviations of informative priors are scaled by factors $.5$, $1.0$, $1.5$, $3.0$, and $6.0$.}
    \label{fig:log:posterior}
\end{figure}

\begin{figure}[htbp]
    \centering
    \includegraphics[width=.7\linewidth]{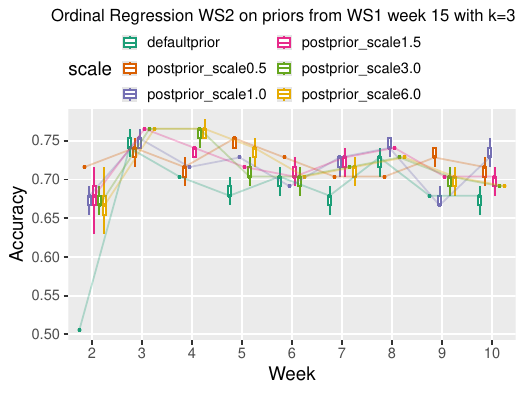}
    \caption{Accuracy of the ordinal regression on the target cohort WS2 with the best prior from WS1 (see \Cref{fig:ord:source}), and with non-informative default prior. Standard deviations of informative priors are scaled by factors $.5$, $1.0$, $1.5$, $3.0$, and $6.0$.}
    \label{fig:ord:posterior}
\end{figure}

Within the range of 10 weeks before Christmas, the following informative priors with scaled standard deviations ($SD$) performed best for the respective model types: %\textcolor{red}{in this itemlist, we should specifically address the results of early weeks and the benefit attained from using prior information. As secondary information, we could discuss how it evolves in later weeks and how the 'best' prior scale evolves over time. I think this is currently reversed.}
\begin{itemize}
    \item \textbf{Linear regression:} The best predictive performance was achieved in week~3, using priors with $6 \times SD$ (see \Cref{fig:lin:posterior}) with an RMSE of $13.17$.
    \item \textbf{Logistic regression:} The best predictive sensitivity was observed at $.83$
    %$.8293$
    in weeks~8 and 9, using priors with $.5\times SD$. The prediction accuracy reached its peak $.74$
    %$.7407$
    in weeks~3 and 5, using priors with $1-1.5$ resp.\ $.5 \times SD$ (see \Cref{fig:log:posterior}).
    \item \textbf{Ordinal regression:} The best predictive accuracy of $.77$
    %$.7654$
    was observed in weeks~3 and 4, using priors with $1.5-6$ resp.\ $3-6 \times SD$ (see \Cref{fig:ord:posterior}).
\end{itemize}

However, the aim in the target cohort WS2 is not merely to find the best prediction across all weeks, but rather to find the earliest possible week, where the predictive performance is already sufficiently high, and to assess whether and to what extent especially these early predictions benefit from the added prior information over the plain Bayesian analysis of current cohort data without informative priors.

%\subsubsection{Predictive Gains from Informative Priors}
For linear regression, the default model performed best already in week~3, whereas its performance deteriorated after week~6. Moreover, we could observe no benefit of incorporating prior information. On the contrary, the RMSE increases the more informative prior information, i.e., with smaller $SD$, was introduced (see \Cref{fig:lin:posterior}).

In contrast, logistic regression benefited from prior information substantially in almost all weeks, with only a few exceptions (see \Cref{fig:log:posterior}). Accurate early prediction of final student performance is possible as early as week~3 or week~4, achieving an accuracy close to $.75$ and sensitivity above $.8$.
For the sake of quantifying the predictive gain from incorporating informative priors, we report the percentage by which the misclassification rate (defined as $1$-ACC), and the false negative rate (defined as $1$-SENS) decreased when prior information derived from WS1 is included in WS2. The results for logistic regression are shown in \Cref{reg:log:mediandecrease}. In week~3 the misclassification rate could be reduced by $22.22\%$, and the false negative rate could be reduced by $38.46\%$ over the uninformative default case, by incorporating informative prior information with their standard deviation scaled by a factor $1-1.5$. However, \Cref{reg:log:mediandecrease} also shows that informative priors did not consistently improve performance across all weeks and prior scales. In later weeks---particularly weeks~9 and 10---several prior specifications increased the misclassification rate relative to the default prior, suggesting that the benefit of prior information diminishes as more current-cohort data accumulate.

\begin{table*}[ht]
\centering
\caption{Improvement of median misclassification and false negative rate in percent (WS2 logistic regression; priors from WS1 week~13). {Largest improvements in bold}. 
}
\centering
\begin{tabular}[t]{lcrrrrrrrrr}
\toprule
& PostPrior & \multicolumn{9}{c}{Weeks}\\
 & Scale & 2 & 3 & 4 & 5 & 6 & 7 & 8 & 9 & 10\\
\midrule
\multicolumn{4}{l}{Misclassification rate} \\ \cmidrule{1-3}
  & .5 & 0.00 & 7.41 & 7.69 & 12.50 & 7.14 & -12.00 & 0.00 & 7.69 & -21.74\\
  & 1.0 & 0.00 & \textbf{22.22} & 3.85 & 4.17 & 10.71 & -4.00 & 3.70 & -11.54 & -21.74\\
  & 1.5 & 0.00 & \textbf{22.22} & 0.00 & -12.50 & 14.29 & -8.00 & 3.70 & -11.54 & -17.39\\
  & 3.0 & -2.94 & 11.11 & -3.85 & -8.33 & 3.57 & -4.00 & 0.00 & 0.00 & 4.35\\
  & 6.0 & -3.03 & 0.00 & -3.85 & -4.17 & 0.00 & 0.00 & 0.005 & 0.00 & -4.35\\
\midrule
\multicolumn{4}{l}{False negative rate} \\ \cmidrule{1-3}
 & .5 & 14.29 & 23.08 & \textbf{38.46} & 33.33 & 33.33 & 20.00 & 36.36 & 36.36 & 0.00\\
  & 1.0 & 5.00 & \textbf{38.46} & 23.08 & 25.00 & 16.67 & 20.00 & 27.27 & 18.18 & 11.11\\
  & 1.5 & 9.52 & \textbf{38.46} & 15.38 & 8.33 & 25.00 & 10.00 & 27.27 & 9.09 & 11.11\\
  & 3.0 & 9.52 & 23.08 & 7.69 & 8.33 & 16.67 & 0.00 & 0.00 & 9.09 & 11.11\\
  & 6.0 & 9.52 & 0.00 & 7.69 & 8.33 & 0.00 & 0.00 & 0.00 & 9.09 & 0.00\\
\bottomrule
\end{tabular}
\label{reg:log:mediandecrease}
\end{table*}

For ordinal regression, prior information tended to improve performance particularly in earlier weeks, while in later weeks performance was often similar or very rarely slightly inferior compared with models estimated without informative priors (see \Cref{fig:ord:posterior}).
Similar to the case of logistic regression, the performance gain results are shown in \Cref{reg:ord:mediandecrease}.

\begin{table*}[htb]
\centering
\caption{Improvement of median misclassification rate in percent (WS2 ordinal regression; priors from WS1 week~15). {Largest improvements in bold.} Remarkable improvement, at higher accuracy (see \Cref{fig:ord:posterior}), in italic.}
\centering
\begin{tabular}[t]{crrrrrrrrr}
\toprule
PostPrior & \multicolumn{9}{c}{Weeks}\\
Scale & 2 & 3 & 4 & 5 & 6 & 7 & 8 & 9 & 10\\
\midrule
.5 & \textbf{42.50} & -4.76 & 0.00 & 20.00 & 8.33 & 7.69 & -9.09 & 15.38 & 11.54\\
1.0 & 35.00 & 0.00 & 4.17 & 15.38 & -4.17 & 15.38 & 4.76 & -3.85 & 18.52\\
1.5 & 35.00 & 9.52 & 12.50 & 11.54 & 0.00 & 14.81 & 4.55 & 7.69 & 7.69\\
3.0 & 32.50 & 9.52 & \textit{20.83} & 7.69 & 0.00 & 11.54 & 0.00 & 7.41 & 3.85\\
6.0 & 32.50 & 9.52 & \textit{20.83} & 16.00 & 0.00 & 11.54 & 0.00 & 7.69 & 3.85\\
\bottomrule
\end{tabular}
\label{reg:ord:mediandecrease}
\end{table*}

It is particularly remarkable, that as early as in week~2, the misclassification rate of the default model---$.51$, which merely constitutes a random guess---could be improved by incorporating informative priors by $32.5\%$ ($3-6\times SD$), $35\%$ ($1-1.5\times SD$), and by $42.5\%$ ($.5\times SD$) to achieve an accuracy of $.72$.
The accuracy of ordinal regression models with weakly informative priors ($3-6\times SD$) peaked at $.77$ %$0.7654$ 
as early as in week~4, where the misclassification was still improved by $20.83\%$ (see \Cref{fig:ord:posterior} and \Cref{reg:ord:mediandecrease}).

Since ordinal regression is a strict generalization of logistic regression, it is interesting to compare their performance when the ordinal predictions are collapsed to a binary classification using the same threshold of 3.3. Ordinal regression estimates the full grade structure through ordered threshold parameters, which may yield a better-calibrated predictor than a model trained only on the binary split, because the ordinal model leverages distinctions among all grade levels to estimate the relationship between engagement features and the latent performance dimension. 

\begin{figure}[htbp]
    \centering
    \includegraphics[width=.65\linewidth]{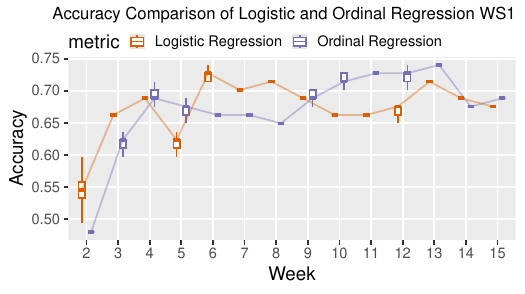}
    \caption{Comparison of prediction accuracy of logistic and ordinal regression on the source cohort WS1.}
    \label{fig:logord:compare}
\end{figure}

\Cref{fig:logord:compare} shows a comparison of the resulting accuracies when applied to the source cohort WS1 where the performances are similar, though the overall picture changes from week to week. We see that logistic regression performs slightly better in earlier weeks on this binary task, which notably constitutes its genuine purpose. However, we stress that ordinal regression provides additional grade information, which can help human instructors tailor individualized decisions and measures to support students who are struggling. 

\begin{figure}[htbp]
    \centering
    \includegraphics[width=.65\linewidth]{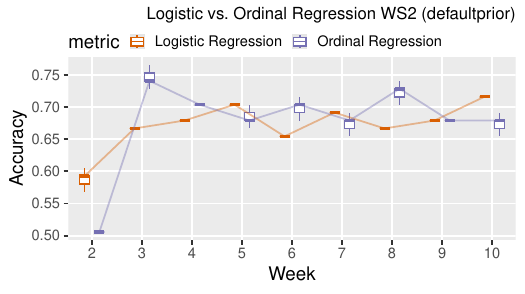}
    \caption{Comparison of prediction accuracy of logistic and ordinal regression on the target cohort WS2 with non-informative default prior.
    }
    \label{fig:logord2:comparedef}
\end{figure}

\begin{figure}[!htbp]
    \centering
    \includegraphics[width=.65\linewidth]{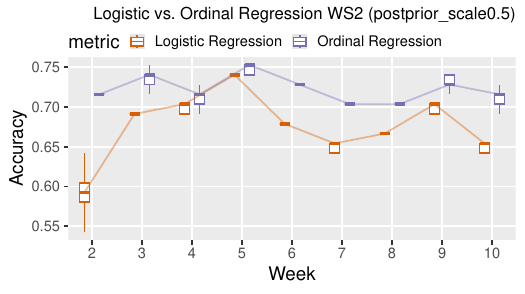}
    \caption{Comparison of prediction accuracy of logistic and ordinal regression on the target cohort WS2 with the most informative prior from WS1, whose standard deviation is scaled by a factor of $.5$.
    }
    \label{fig:logord2:comparep05}
\end{figure}

We next performed the comparison on the target cohort. In the case that the default prior is used in WS2, logistic and ordinal regression again yield comparable results (see \Cref{fig:logord2:comparedef}) as before in WS1. The most compelling advantage of ordinal regression becomes apparent when we incorporate the informative prior information from WS1 into the WS2 model: this puts prediction accuracy based on the ordinal regression clearly on top, with accuracies always better than $.7$ (see \Cref{fig:logord2:comparep05}). It also underlines that ordinal regression benefits more strongly from the added prior information than logistic regression. This can be seen especially in week~2, where logistic regression remained at an accuracy of $.59$, %$.5926$ 
while ordinal regression improved from $.51$ to $.72$ (cf. \Cref{fig:ord:posterior}). %$.5062$ to $.7160$.

\subsubsection{How Prior Information Shapes Posterior Parameter Estimates}

\begin{figure}[htbp]
    \centering
    \includegraphics[width=.97\linewidth]{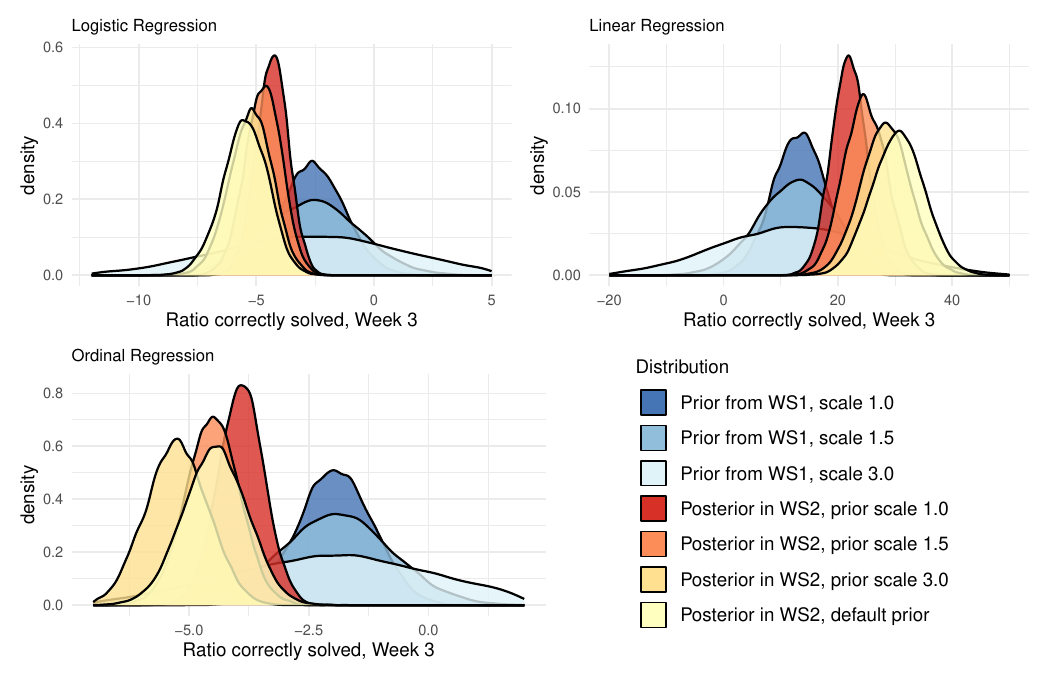}
    \caption{Kernel density plots of the coefficient distributions for the variable ``Ratio of correctly solved problems in the online exercise in week~3'' in the linear, logistic and ordinal regression models.}
    \label{fig:density_illustration_comparison}
\end{figure}

To illustrate the effect of incorporating more or less informative prior parameter distributions, we inspect the exemplary variable ``Ratio of correctly solved problems within the online exercise in week~3''. \Cref{fig:density_illustration_comparison} shows kernel density estimates for all three regression types. In each panel, shades of white to blue represent the prior distributions derived from WS1 at different scaling strengths, and shades of light-yellow to red represent the resulting posterior distributions in WS2---ranging from the uninformative default prior (lightest) to the most informative prior specification (darkest).

The figure shows that as prior strength increases (i.e., as the scaling factor decreases), the resulting posterior distributions become more concentrated and shift more strongly toward the prior mean. Under the uninformative default prior, the posterior is driven almost entirely by the current-cohort likelihood, whereas stronger priors pull the posterior noticeably toward the estimates obtained in WS1.

\begin{figure}[htbp]
    \centering
    \includegraphics[width=.5\linewidth]{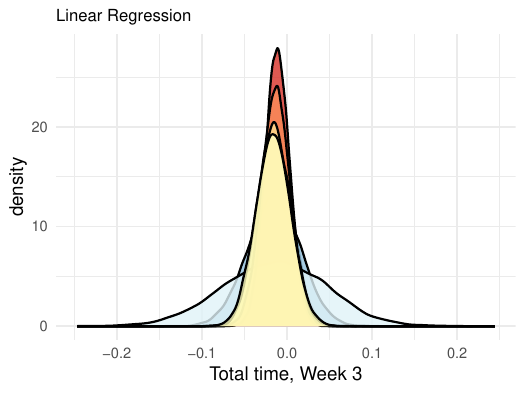}
    \caption{Kernel density plot of the coefficient distribution for the variable ``Total time spent working on the online exercise sheet in week~3'' in the linear regression model. The color encoding equals that of \Cref{fig:density_illustration_comparison}.}
    \label{fig:density_linreg_noeffect}
\end{figure}

Examining the parameter distributions for the variable that measures ``Total time spent working on the online exercise sheet in week~3'', explains why the numerical linear regression models did not benefit as much as the other regression models for classification. As can be seen in \Cref{fig:density_linreg_noeffect}, all prior as well as posterior parameter distributions overlap almost congruently differing only slightly in terms of the scale of their $SD$. Thus, incorporating prior information widened or narrowed the resulting posterior distributions very little while leaving their location virtually unchanged, which means that on average all models produced very similar predictions.

\section{Discussion}

The proposed Bayesian updating procedure performed effectively as a method for developing an early-warning system, especially for identifying students at risk. Incorporating prior information from the earlier cohort often led to considerable improvements in predictive performance. Employing linear models to predict the exam scores of the final exam led to predictions within $\pm 13.15$ points around the actual score. However, priors and posterior distributions were so similar for linear regressions that the models did not benefit and even degraded, when priors and likelihoods disagreed.

In contrast, logistic and ordinal regression models for classification were not only very accurate by week~6, but including informative priors from the previous cohort improved the models to almost full predictive performance as early as in weeks~2 or 3, where the default model that only used information from the currently running semester was not sufficient to achieve satisfactory prediction accuracy or sensitivity. The performance improvements measured by the percentage by which misclassifications were reduced, reached $20.83\%$ up to even $42.5\%$ especially in these early weeks, and the rate by which students at risk were not identified by the logistic regression model could be reduced by $38.46\%$ in the early week~3.

The results suggest that using informative priors to flexibly augment currently observed data is beneficial and selecting prior strength based on scaling standard deviations from the previous cohort provides a practical and effective approach for fine-tuning.

Comparable early performance levels have typically required personal demographic, or self-report data (see Arizmendi et al., \citeyear{Arizmendi.etal2022}, for an overview), or data from prior-knowledge exams \citep[see, for example,][]{cogliano2022self}, that are often not available or incomplete, or pose risk of stigmatization. \citet{Bernacki.etal2020}, for example, achieved comparable sensitivity rates using behavior-only LMS data in a different educational context. Similar to this work, our models relied only on malleable behavioral variables from course LMS logs. However in contrast, we used only variables derived from logs of two practice opportunities and no other sources. Beyond the choice of predictors, the Bayesian updating approach offers a privacy advantage: prior cohort information enters the model as a parameter distribution rather than as individual student records, so no personal data from previous cohorts need to be retained or transferred.

\subsection{Contribution to the Literature}

This study extends current research in learning analytics and educational data science in three key ways. First, our findings reinforce the value of using malleable, SRL-aligned behavioral indicators for early prediction of student performance. Prior work has highlighted the limitations of models built on static or demographic variables, as they offer little pedagogical value for intervention \citep{Arizmendi.etal2022}. 
By prioritizing changeable learning events---such as the ratio of correctly solved problems---over non-malleable factors like socioeconomic status, our model shifted the goal of learning analytics from passive prediction to proactive support \citep{Bernacki.etal2020}. Following the prediction-and-intervention design \citep{cogliano2022self}, identifying at risk students based on these malleable traces provides instructors with clear `lever points' for deploying targeted instructional interventions before students reach graded assessments \citep{Bernacki.etal2020}.
%By relying exclusively on weekly practice- and feedback-related engagement measures, this study confirms that early and actionable risk detection can be achieved using behaviors that students can modify---thereby supporting intervention-focused learning analytics. 
Moreover, the added value of prior cohort information is learned from individual data, but summarized via prior distributions, so that no knowledge of individual's prior data need to be included. 
More specifically, the accuracy of $.77$ achieved by week~4 compares favorably to the mean accuracy of $.72$ reported across $82$ prediction models in a recent review \citep{Arizmendi.etal2022}, particularly given that our models relied exclusively on log data from two learning opportunities: the tutorial videos and the retrieval practice online exercises.

Second, this work contributes to the ongoing discussion about the generalizability and transferability of predictive models in education. Consistent with evidence that model performance declines when applied to new cohorts \citep{Conijn.etal2017, Xing.etal2021}, our results demonstrate that Bayesian updating can meaningfully mitigate this decline. In doing so, the study broadens the empirical base for Bayesian transfer models beyond the predominantly online learning settings examined in prior research.

Third, this study advances the field by providing practical methodological guidance on how to incorporate prior information into predictive student models. While the potential of Bayesian statistics for cumulative knowledge building in education has repeatedly been emphasized \citep{Konig.VanDeSchoot2018}, concrete empirical examples demonstrating how to extract, scale, and apply informative priors remain scarce. We address this gap by systematically comparing prior specifications and showing that informative priors---constructed from posterior distributions of the previous cohort---yield the most reliable predictions. This nuance offers a replicable blueprint for researchers and institutions aiming to build multi-year early-warning systems that improve rather than restart each year.

\subsection{Limitations and Future Research Directions}
The presented work goes beyond the settings that are prevalent in existing literature, where either data is available and examined only for one cohort, or a predictive model obtained from one cohort is directly applied to a second cohort of the same course. 
Our work integrates two consecutive cohorts in that all available information from the first cohort builds a prior information from which the predictive model trained from second cohort data can beneficially be augmented and updated cumulatively. However, this can merely be regarded a first step. It will be an important and necessary future direction to investigate the long-term validity of the presented Bayesian updating methods in a longitudinal study ranging over more cohorts.

In particular, it remains unclear whether the propagation of prior information across several cohorts remains valid and beneficial over time. In a realistic teaching environment, prior information from an earlier cohort may remain useful for a few consecutive iterations, but is likely to lose validity as courses evolve. Changes in course design, demographic shifts across student populations, or a change in course instructors could all alter the relationship between engagement behaviors and performance outcomes in ways that are not captured by priors derived from earlier cohorts — or that actively conflict with them. This would require a mechanism for gradually downweighting older prior information rather than accumulating it indefinitely.

A further limitation concerns the restriction of the analytic sample to students who attended the first exam date. Students who registered only for the second sitting were excluded, and it is plausible that this group is disproportionately at risk. To the extent that this is true, the present models were developed and validated on a sample that underrepresents the most vulnerable students, which may lead to optimistic estimates of real-world early-warning performance. Future work should mitigate this issue by (i) modeling the participants of the two exam dates in separate groups and (ii) integrating these groups into a joint hierarchical Bayesian model, which constitutes a second unique strength of Bayesian models; however, this is beyond the scope of the present study, which focused on iterative and cumulative Bayesian model updates.

Overall, Bayesian modeling opens up intriguing avenues for future research toward synchronizing statistical models from LMS data with the everyday requirements of university courses and instructors by building adaptive models cumulatively and dynamically across cohorts. This would contribute to the overarching goal of a reliable early-warning system that identifies students at risk and supports instructors in deploying targeted interventions to improve student outcomes.

% \textcolor{red}{
% Modell mit 0-Imputation und Jakobs Zusatzfeatures:\\
% - 0-Imputation ist in unserem Fall semantisch vertretbar und hat sich im Vergleich zu ''intelligenteren'' Imputationsverfahren durchgesetzt.\\
% - Zusatzfeatures haben nach hinten raus leicht schlechtere Modelle gebracht, aber die frühzeitige Prädiktion hat sich leicht verbessert; dementsprechend haben wir sie behalten.
% }

\subsection{Open Practices Statement}
Due to data protection rules, the data cannot be made available, but documented code and a simulated dataset based on the empirical data are available at \\ \url{https://osf.io/bquzy/overview?view_only=82d5e70ec17a41cdb52330bddbdb9e3b}.\\
The study was not preregistered. Ethical approval for the original data collection was granted by the ethics committee of the\ifthenelse{\boolean{anonymous}}{
    % Anonymous version: omit author details
    university ([blinded for review])
    %\date{}
}{
University of Hohenheim, Germany,} at which the study was conducted. Participants provided informed consent prior to data collection.

% \section*{Acknowledgments}
% We thank the FAIR project, and some others.

%\subsection*{Contributions}

%Author contributions: \insertcreditsstatement

% {\appendix[Proof of the Zonklar Equations]
% Use $\backslash${\tt{appendix}} if you have a single appendix:
% Do not use $\backslash${\tt{section}} anymore after $\backslash${\tt{appendix}}, only $\backslash${\tt{section*}}.
% If you have multiple appendixes use $\backslash${\tt{appendices}} then use $\backslash${\tt{section}} to start each appendix.
% You must declare a $\backslash${\tt{section}} before using any $\backslash${\tt{subsection}} or using $\backslash${\tt{label}} ($\backslash${\tt{appendices}} by itself
%  starts a section numbered zero.) See the outcommented example below.}


% \section{References Section}
% You can use a bibliography generated by BibTeX as a .bbl file.
%  BibTeX documentation can be easily obtained at:
%  http://mirror.ctan.org/biblio/bibtex/contrib/doc/
%  The IEEEtran BibTeX style support page is:
%  http://www.michaelshell.org/tex/ieeetran/bibtex/
 

%\clearpage

% \bibliographystyle{IEEEtran}
%\bibliographystyle{plain}
%\bibliographystyle{abbrv}
\bibliography{references}

% \newpage

 

% \end{IEEEbiography}



%\vfill

\end{document}